\documentclass{sig-alternate-clc}
\usepackage{graphicx}
\usepackage{subfigure}
\usepackage{url}
\usepackage{numprint}
\usepackage{array}
\usepackage{multirow}

\newcommand{\MyCaption}[3]{\caption[#1]{#1 #2\label{#3}}}
\newcommand{\MyFigureReference}[1]{(see Figure \ref{#1})}
\newcommand{\MyTableReference}[1]{(see~Table~\ref{#1})}
\newcommand{\MyIgnore}[1]{}

\newcommand{\MyFamilySizeMax}[0]{\ensuremath{c_{\mbox{hard}}}}
\newcommand{\MyFamilySizeMin}[0]{\ensuremath{c_{\mbox{soft}}}}
\newcommand{\MyHostCapacity}[0]{\ensuremath{h_{\mbox{cap}}}}
\newcommand{\MyMaxDOs}[0]{\ensuremath{n_{\mbox{max}}}}
\newcommand{\MyMaxHosts}[0]{\ensuremath{h_{\mbox{max}}}}

\newcommand{\MySimulationTime}[0]{\ensuremath{S_{\mbox{t}}}}
\newcommand{\MyHline}[0]{\noalign{\hrule height 2pt}}

\npstyleenglish

\begin{document}

\title{When Should I Make Preservation Copies of Myself?}

\numberofauthors{2}
\author{
\alignauthor Charles L. Cartledge\\
       \affaddr{Department of Computer Science}\\
       \affaddr{Old Dominion University}\\
       \affaddr{Norfolk, VA 23529 USA}\\
       \email{ccartled@cs.odu.edu}\\
\alignauthor Michael L. Nelson\\
       \affaddr{Department of Computer Science}\\
       \affaddr{Old Dominion University}\\
       \affaddr{Norfolk, VA 23529 USA}\\
       \email{mln@cs.odu.edu}\\
}
\maketitle


\begin{abstract}
We investigate how different preservation policies ranging from least aggressive
to Most aggressive affect the level of preservation achieved by
autonomic processes used by smart digital objects (DOs). 
The mechanisms used to support preservation across different hosts
can be used for  automatic link generation and  support preservation
activities by moving data preservation from an archive centric perspective
to a data centric preservation.
Based on simulations of
small-world graphs of DOs created using the Unsupervised Small-World algorithm, we
report quantitative and qualitative results for graphs ranging in size from 10 to
\numprint{5000} DOs.  Our results show that a Most aggressive preservation policy makes
the best use of distributed host resources
while using one half of the number of messages of a Moderately
aggressive preservation policy.
\end{abstract}


\section{Motivation}
Much of our current cultural heritage exists only in digital format and
digital preservation approaches rely on the long term commitment of individuals,
institutions and companies to preserve this heritage.  The length of time that
an individual will be engaged in preservation activities is, by definition, limited
to their lifetime (and probably just the middle part of that life).  Even those few years
may be longer than institutions and companies would be willing to undertake
digital preservation. Institutions and companies may cease to
exist or be unwilling or unable to meet their original preservation  commitments due to changes
in corporate culture or financial considerations.   If this
happens then the digital files and their information (our heritage) may become
irretrievably lost.  The acknowledgement that much of our 
 heritage exists only in digital format, and the recognition
that there is a real risk of total loss through accident \cite{yin2011flickr} or change in 
business goals \cite{milian2009geocities}
has been recognized in academic reports
and papers \cite{1555440} and
is starting to surface
in the popular press \cite{walker2011CyberDead,wohisen2011NeverDies,smith2011friendster,gross2012megaupload}.

Our motivation is to change the focus from preservation services
administered by institutions (a repository-centric perspective) to
one where the data preserves itself (a data-centric perspective). We
continue to investigate this data-centric perspective through the use of
the Unsupervised Small-World (USW) graph creation algorithm
\cite{jcdl09-usw,1378990,cartledgeTechreport,cartledge2010analysis}
where we have shown that DOs
instrumented with just a few rules can autonomously form
into small-world graphs.   The focus of this work
is to augment the prior work 
by imbuing  DOs with the capability
 to create a number of copies
of themselves for preservation purposes.
We are focusing on determining when  copies
should be created during the USW process and  the communication
impacts  of different preservation policies.

\section{Related Work}\label{sec:related_work}
This work is at the convergence of
digital library repositories, emergent behavior, 
graph theory and web infrastructure.
To provide a context for understanding
the contributions of this research, we
first briefly review the status of how
objects are stored in repositories as well
as the nature and types of various networks
or graphs.

\subsection{Repositories}

Repositories  range from theoretical to
ready-to-download.  Some such as SAV \cite{699622} are frameworks or architectural proposals.  
Some, like FEDORA
\cite{700063}, are middle-ware systems, ready to be the core repository
technology in a local deployment.  Some such as aDORe \cite{adore:tcj} are complete systems, ready
to deploy.  These include DSpace \cite{700078}, sponsored by
MIT and HP Laboratories and LOCKSS \cite{1047917}, sponsored by the
Stanford University Libraries.  All are widely implemented and enjoy a
large user community.  DSpace is an institutional repository, intended to
archive the intellectual output of a university's faculty and students.
LOCKSS allows libraries to create ``dark archives'' of publishers'
websites.  As long as the publishers' websites are available, all web
traffic goes to those sites.  But if the publishers' contents are lost,
the dark archives are activated and the content is available again.
Risk is mitigated through many sites archiving content of their own
choosing.  Depending on an institution's requirements, the systems
described above can be quite attractive.  But there is an implicit
assumption on any repository system:  that there is a person, community
or institution that exists to tend to the repository.  \emph{What happens when
the responsible organization no longer exists?}  There are repository
trading and synchronization provisions (e.g., \cite{506310}), but most
are specific to a particular repository architecture.

Cooperative
File Systems (CFS) \cite{cfs:sosp01}, Internet Backplane Protocol (IBP)
\cite{633058}, Storage Resource Broker (SRB) \cite{823379} and OceanStore
\cite{613636} are among  several generic network
storage systems and APIs that have also been proposed.
CFS and OceanStore rely on distributed
hash tables and an overlay network to locate content in the Internet.
Systems with such additional levels of shared infrastructure have not been widely
deployed.  IBP and
SRB are more traditional in their repository design and have enjoyed
greater deployment.  SRB (and its follow-on, iRODs \cite{rajasekar2006prototype})
 has
a user community similar in size to LOCKSS and Fedora.

Numerous P2P repositories have also been proposed (for example Intermemory \cite{785953}, Freenet \cite{371977},
Free Haven \cite{371978}, and PAST \cite{502053}).  Frequently
these repositories are characterized by offering long-term storage systems requiring the
contribution of X megabytes of storage today for the promise of
Y megabytes of persistent storage (X $\gg$ Y).
 Despite having
many theoretically attractive features, these systems have not found
widespread acceptance.  We use a variant of this idea in our graph construction 
techniques, by simulating that a host has effectively infinite capacity 
for those DOs that are created locally and a very limited 
capacity for those DOs that were created remotely.

Each of the approaches listed above inherently rely on human and institution intervention in
the digital preservation activities of refreshing and migration \cite{waters1996pdi,rothenberg:atq}.
The digital preservation activities of emulation and metadata attachment are outside
the context of this paper.
 As the amount of digital data continues
to grow (at potentially an exponential rate), the organizational and human
cost to keep up with traditional approaches will become overwhelming.  An alternative
approach is to revisit the definition of a DO and to incorporate into that
definition the idea that the DO is empowered to make preservation copies of
itself for the purposes of preservation and that it can communicate with other DOs.
Messages that can be sent include the location of new supporting preservation hosts, 
data migration services and new DOs.

\subsection{Graph Construction}

Our approach for the construction of a small-world network of DOs for self preservation is
different than others have used or proposed. We  make use of the definition of a small-world graph
as one that has a high \emph{clustering coefficient} when compared to a 
randomly created graph and an average path length that is proportional to the
number of nodes in the graph \cite{watts:collective_dynamics}.
The Watts-Strogatz approach to constructing such a graph is to take a lattice graph of  degree $k$ and size $n$ and
perturb the links to create a graph with small-world characteristics.
Some approaches make connections
between nodes  based on the proportion of the destination node's degree count
\cite{nguyen:analyzing_and_characterizing_small_world_graphs,klemm2002gsf,barabsi:world_wide_web_forms_a_large_directed_graph},
 a kind of preferential
attachment or fitness policy.  
Yet another type of approach takes an existing graph and then grows a small-world by the addition of new links
\cite{duchon:towards_small_world_emergence,kleinberg:small_world}.
Or, by connecting a node to a fixed number of vertices based on
their degree  \cite{bollobas2001dss}, or even creating a small-world
graph from a random one \cite{gaume:from-random-graph}.

The USW process requires that each new node communicate with an existing 
node in the USW graph.
After the first DO selection, the USW algorithm controls
where the DO fits into the graph and how many edges are created
to other DOs in the system.  USW is the only small-world
graph creation algorithm that we know of where connections
are made between DOs based information that the  DO
gleans prior to making its first connection.

\section{Self-Preserving Digital\\Objects}

We consider DOs to be in the tradition of
Kahn-Wilensky and related implementations \cite{kahn2006fdd}.
This paper focuses on the analysis of inter- and intra-DO policies for 
preservation through simulation.  In a separate project we are 
implementing a test bed of DOs as web resources that use OAI-ORE \cite{ore:techreport:2008} 
Resource Maps to keep track of the contents of DOs, the location of 
supporting web services, and the JavaScript necessary to implement the 
policies presented here.  The test bed will feature DOs that utilize a 
variety of storage layers, such as repository systems (e.g., DSpace, 
FEDORA), file systems, web storage services (e.g., Amazon S3), wikis, 
blogs, and email accounts (e.g., Gmail).

\MyIgnore{
A collection of DOs can be viewed  from a graph theoretical perspective 
as made up of vertices and
undirected edges.
Each DO contains the identities of DOs that are in
its \textit{k}-neighborhood (\textit{k} = 1) and
each DO uncovers  the identities of its \textit{k}-neighborhood through a discovery process.
The USW graph structure has the properties of a small-world graph but is created via autonomic processes
vice perturbing an existing lattice.}

\subsection{Flocking for Preservation}\label{sec:flocking}

Craig Reynolds' seminal paper on ``boids'' 
\cite{reynolds:flock},  demonstrated that three simple rules
were sufficient to simulate the complex behaviors of schools of fish,
flocks of birds, herds of animals and the like.  
The remarkable feature about these rules is that they are scale-free so
knowing the
entire size of the group, or network is not required.  We believe these
rules can be adapted to create self-preserving DOs with
similarly complex emergent behaviors.  \MyIgnore{Table \ref{flocking} lists the
rules that Reynolds proposed for boids
and our interpretation within the DO preservation world.}
The transcription of
Reynolds' rules from a boid to a DO perspective are:

\textit{Collision avoidance}  DOs flocking to a new repository
cannot overwrite each other (collide in physical storage), nor collide
in namespaces (have the same URI). This is orthogonal to the naming
mechanism used: URIs, URN  handles,
DOIs, globally unique identifiers (GUIDS)
or content addressable naming schemes \cite{383072}.

\textit{Velocity matching} All members of a herd, or school, or flock
move at roughly the same speed. With boids, the idea  is to travel the same
speed as your neighbors. Interpreting  velocity  as resource consumption
(i.e., storage space) enables this rule to be applied to a DO environment. Specifically,
a DO should try to consume as much, and only as much, storage as
everyone else. In resource-rich environments (lots of storage space
available on lots of hosts), making as many copies
of yourself as you would like is easy. When storage becomes scarce,
this becomes more difficult. DOs must be able to
delete copies of themselves from different repositories to make room for late
arriving DOs in low-storage situations. DOs will never delete
the last copy of themselves to make room for new DOs, but they will
delete copies of themselves to come down from a soft threshold (e.g.,
10 copies) down to a hard threshold (e.g., 3). When resources become
plentiful again, new copies can be made.

\textit{Flock centering} For boids this  means staying near (but not
colliding with) other flock-mates.  We interpret this in a manner similar
to \textit{velocity matching}, with
DOs attempting to stay near other DOs as they make copies of
themselves at new repositories. In essence, when a DO learns of
a new repository and makes a copy of itself there, it should tell the
other DOs it knows so they will have the opportunity to make copies
of themselves at the new location. Announcing the location
of a new repository will thus cause DOs at other repositories that
have not reached their soft threshold to  create copies that ``flow'' to the
new repository.  

The benefits of using the boids model are: it is simple to
implement and test (cf., iRODs); all decisions are made using locally
gleaned information; there are no global controls with the attendant
communications overhead costs; and once a DO is created and introduced into the USW
network, the DO is responsible for its destiny.   Simple rules that
are executed based on locally gleaned information result in emergent \emph{intelligent} 
and \emph{social} behaviors.

At the macro level; in much the same  way that flocks self-navigate to new locations that have
the resources they need, we envision DOs
self-preserving in a  loose confederation of cooperating repositories each
with varying levels of resources and availability.  Making copies
in new repositories is performed in an opportunistic manner,
within the guidelines imbued in the DOs at creation time.  From time to time
an archivist may steer the
entire collection (or parts of it) to new archives, but for the most part the DOs replicate
and preserve themselves.

\begin{sloppypar}
\subsection{Unsupervised Small-World Graph Creation}
\end{sloppypar}

We introduce some terminology to discuss how DOs can self-arrange.
\emph{Friends} are DOs that share an edge.  When a DO is created, is introduced to an existing DO in the graph and is called a \emph{wandering} DO.
While wandering,  a DO
accumulates a list of potential friends from other DOs in the graph.
When a wandering DO makes its first friendship link
to a DO, the no-longer wandering DO uses the information
that it has gleaned about other DOs  to create additional friendship
links.   This process with 4 DOs is shown in Figure \ref{fig:overview}.
Friendship links are separate from HTML navigation links (i.e., \texttt{<link>}
instead of \texttt{<a>} HTML elements).  
 A \emph{family} is the collection of
 DOs that are replicas of each other.  A \emph{parent} is the family
member  responsible for meeting the family's
 preservation
goals.

\begin{figure}
\centering
\newcommand{\MySubfigure}[3]{\subfigure[#1]{\includegraphics[trim = 10mm 35mm 10mm 35mm, clip,  width = 1.0in] {overviewStage00#3.ps.eps}}}
\newcommand{\MySubfigureA}[3]{\subfigure[#1]{\includegraphics[trim = 10mm 160mm 10mm 35mm, clip, width = 1.0in] {overviewStage00#3.ps.eps}}}
\MySubfigureA{DO$_{1,0,*}$}{ignoreReference}{1}
\MySubfigureA{DO$_{2,0,*}$ contact}{ignoreReference}{2}
\MySubfigureA{DO$_{2,0,*}$ link}{ignoreReference}{3}
\\
\MySubfigure{DO$_{3,0,*}$ first contact}{ignoreReference}{4}
\MySubfigure{DO$_{3,0,*}$ second contact}{ignoreReference}{5}
\MySubfigure{DO$_{3,0,*}$ first link}{ignoreReference}{6}
\\
\MySubfigure{DO$_{4,0,*}$ first contact}{ignoreReference}{7}
\MySubfigure{DO$_{4,0,*}$ second contact}{ignoreReference}{8}
\MySubfigure{DO$_{4,0,*}$ first link}{ignoreReference}{9}
\MyCaption{The USW growth algorithm with 4 DOs.}
{The ``wandering'' DO symbol is filled.   Dashed lines are communications.  
Solid lines are friendship links.}
{fig:overview}
\end{figure}

Friendship
links serve as a way for DOs to send messages from one to another,
such as when new storage locations are available or
the scope and migration of file formats (cf. the semi-automated alert
system described in Panic \cite{996415}).  Friendship links are used to
support the preservation process and  meet the 
spirit of preservation refreshing.

\begin{sloppypar}
\section{Self-Preservation Policies for Preservation}
\end{sloppypar}
\subsection{Model}
We simulate  three different replication policies to quantify and qualify 
their effects on the system as measured in two different
areas.  The first area being how effective the replication policy
 is at having as many DOs as possible achieve their desired
maximum number of preservation
 copies.
  The second being the communication costs associated with each
replication policy as the system grows in size.

A DO's family members will be spread across a collection of hosts.  A complete
description of a DO's position in a family structure and the host
that it is living on is given by the notation DO$_{n,c,h}$.  Where:\\
\ensuremath{
\begin{array}{l l}
&\MyFamilySizeMin = \mbox{min. preservation copies}\\
limits:&\MyFamilySizeMax = \mbox{max. preservation copies}\\
&\MyMaxDOs = \mbox{max. DOs}\\
&\MyMaxHosts = \mbox{max. hosts}\\
\\
n,c,h\mbox{ defined as:} &
\begin{array}{l}
 n = 1, \ldots, \MyMaxDOs \\
 c = 0, \ldots, \MyFamilySizeMax \\
 h = 1, \ldots, \MyMaxHosts \\
\end{array}
\\
\\
\mbox{subject to:} &
\begin{array}{l l}
(n,h) \mbox{ unique } \forall \mbox{ n and } \forall \mbox{ h} \\
c=  \left\{\begin{array}{l l}
0 & \mbox{if parent DO} \\
>0 & \mbox{otherwise} \\
\end{array}
\right. \\
\end{array}
\end{array}
}

If $c >0$ then $c \leq \MyFamilySizeMin \leq \MyFamilySizeMax$.

\subsection{Policies}
We focus on the following preservation policies (assuming that the DO values for 
\MyFamilySizeMin~and \MyFamilySizeMax~ have been defined):
\begin{enumerate}
\item \textbf{Least aggressive} --- a DO will  make only a single preservation copy at a time, regardless of how
many copies are needed, or how many opportunities are available  and will continue
to make single copies until it reaches \MyFamilySizeMax.
\item \textbf{Moderately aggressive} --- a DO will make as many copies as it can to reach \MyFamilySizeMin~
when it makes its first connection, then  fall back to \emph{Least aggressive} policy.
\item \textbf{Most aggressive} ---  a DO will make as many copies as it can to reach \MyFamilySizeMax~
when it makes its first connection, then  fall back to \emph{Least aggressive} policy.
\end{enumerate}

The effect of both the Moderately and Most aggressive preservation behaviors is that after reaching their respective
goals, they behave like the Least aggressive.

\subsection{Evaluation}
\begin{figure}
\centering
\includegraphics[width = 3.5in]{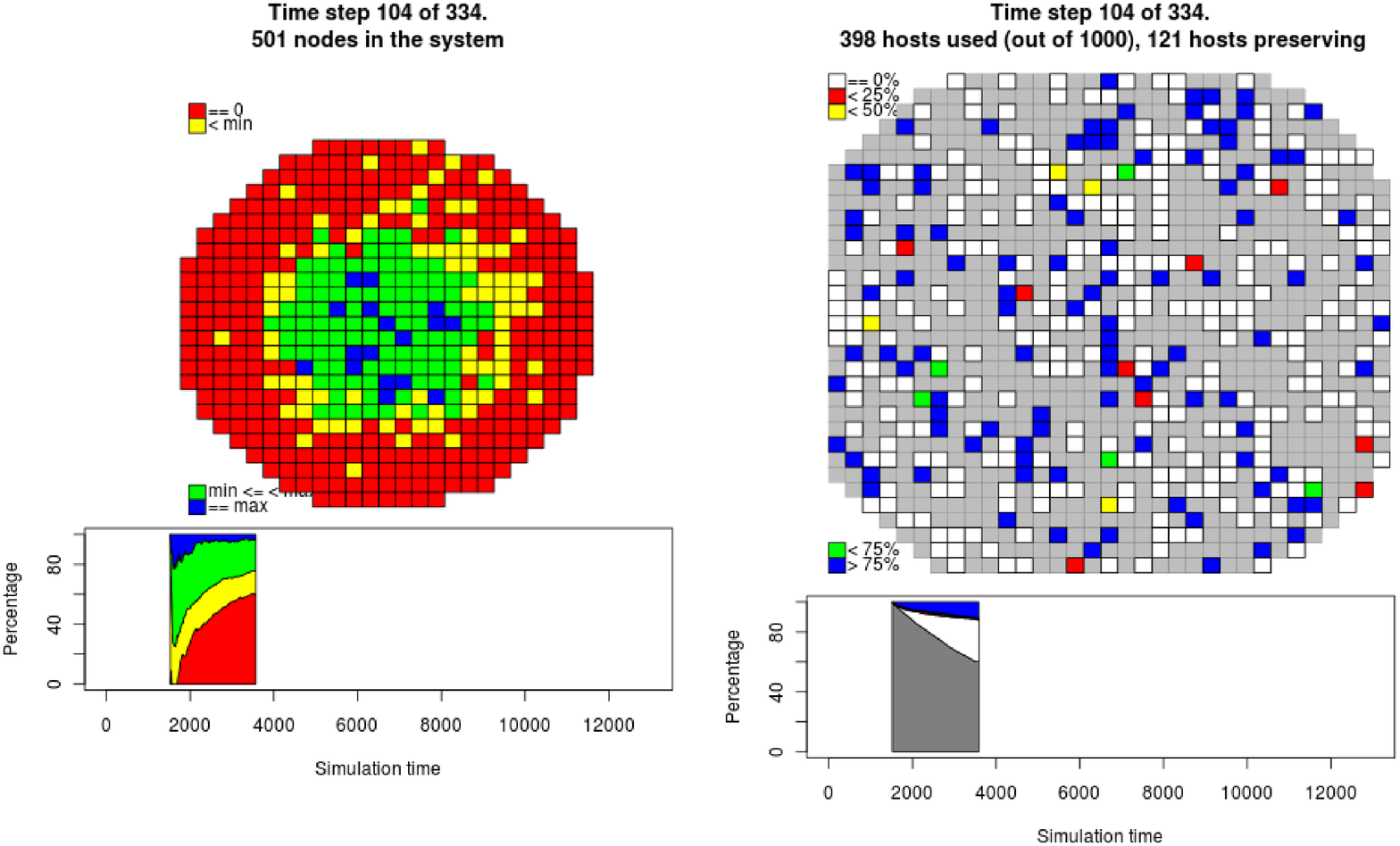}
\MyCaption{A snapshot of the Least aggressive preservation policy.}
{DOs are shown on the left and hosts are shown on the right.   The colors show the state of the DO's preservation copies, 
 or host's preservation capacity used at the time of the measurement. Under each circular plot is a \MySimulationTime histogram. Above
each circular plot is a status line showing \MySimulationTime, how many DOs are in the system or how many hosts are active and preserving data.}
{fig:legend}
\end{figure}

Figure \ref{fig:legend} serves as a legend for the sub-figures in Figures \ref{fig:timestep:legend} and  \ref{fig:1} and shows DO and host preservation status
as a function of \MySimulationTime.   Figure \ref{fig:legend} is divided into four areas.
The left half  shows DO related data, while the right half shows host data.  DOs are  sequentially
 added to the simulation.  In Figure \ref{fig:legend}, DOs are added in a spiral fashion starting at the center of the ``circular'' plot, with
newer DOs are plotted in a circular manner from the center.  This presentation is similar to the rings of a tree,
 the oldest are in center and the youngest are on the outer edge.

The preservation status of a DO is approximated by the color assigned to the DO.  Initially the DO has $c=0$ copies and is colored
red.  As the DO creates copies, the color changes to yellow.  When the DO reaches \MyFamilySizeMin, the color
changes to green.  When \MyFamilySizeMax~ is reached, the DO turns blue.  The  rules of the simulation (based on our
interpretation of Reynolds' ``boids'') permit the killing of one DO's preservation copies for the sake of creating
room for copy of a DO that needs to reach its \MyFamilySizeMin~ (i.e., if a DO$_{i,c,h}$ has more than its \MyFamilySizeMin~ and DO$_{j,c,h}$ has not
reached its \MyFamilySizeMin, then DO$_{i,c,h}$ will sacrifice one of its copies so that DO$_{j,c,h}$ can move closer to
\MyFamilySizeMin).  Sacrificing a preservation copy for the betterment of the whole is the embodiment of \textit{velocity matching}.
The effect of this behavior is that a DO can change color from red to yellow to green and then possibly to blue.  If the
DO changes to blue, it might oscillate between green and blue as its number of preservation copies oscillate between \MyFamilySizeMin~ and
\MyFamilySizeMax.  A DO will never
sacrifice a copy if it has not exceeded its \MyFamilySizeMin.  The  histogram under the DO circular plot shows the
percentage of DOs in each of the different preservation copy states as a function of \MySimulationTime.

The preservation utilization status of a host is shown in the right half of Figure \ref{fig:legend}.  The universe of possible
hosts is constant and is represented by the entire right half plot.  Hosts that are not being used are shown in grey.  The placement
of the host in the figure is based on the host's sequential number in the simulation.
Those hosts that are used are drawn in one of five colors.  If the host is used in the simulation, but is not hosting any preservation copies
then it is colored white.  If less than 25\% of the host's capacity is used then it is colored red.  Similarly, it is yellow if less than
50\% is used, green if less than 75\% and blue if greater than 75\%.  The histogram on the host's side shows the percentage of the hosts
that are in any of the particular states.

In the simulation, each host has a finite amount of storage that makes available for DOs that originate from other hosts.
This storage is called $\MyHostCapacity$.
The simulation has \MyMaxDOs=500, \MyFamilySizeMin =3, \MyFamilySizeMax = 5, $\MyMaxHosts = 1000$, $\MyHostCapacity = 5$. 
The simulation runs until it reaches a steady state.  A steady state is defined as when the system stops evolving.
Evolution stops when DOs are unable to locate candidate hosts on which to
store additional preservation copies.  Steady state is reached at different times based on the preservation policy.  In all cases, all
$n_{\mbox{max}}$ DOs have been introduced into the simulation by $\MySimulationTime= 3500$.

\begin{figure*}[ht]
\newcommand{\MySubfigure}[3]{\subfigure[#1]{\label{#2}\includegraphics[width = 2.85in] {#3}}}
\centering
\MySubfigure{\MySimulationTime = 1500.}{sub:fig:tl:1}{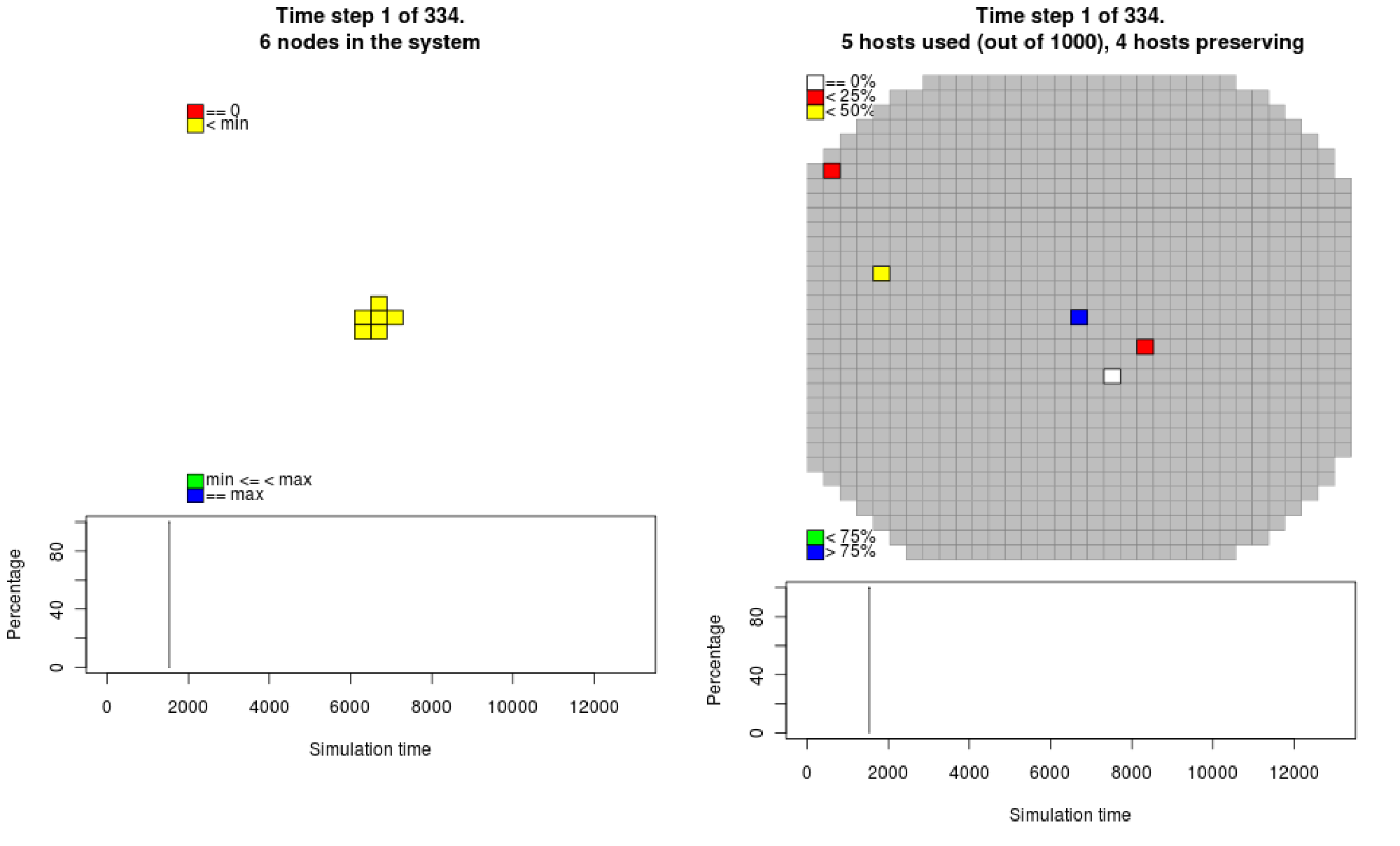}
\MySubfigure{\MySimulationTime = 1700.}{sub:fig:tl:2}{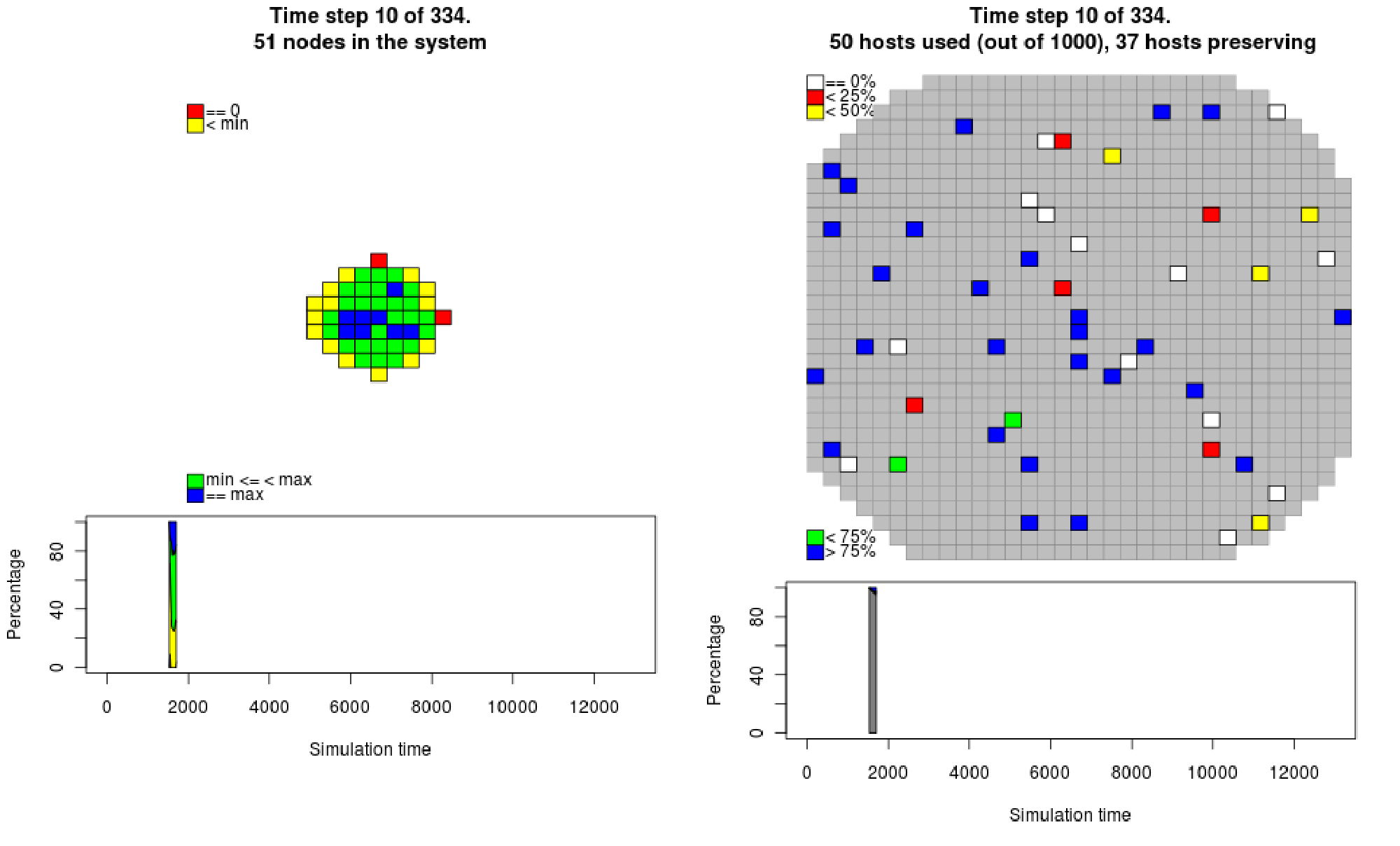}
\\
\MySubfigure{\MySimulationTime = 2200.}{sub:fig:tl:3}{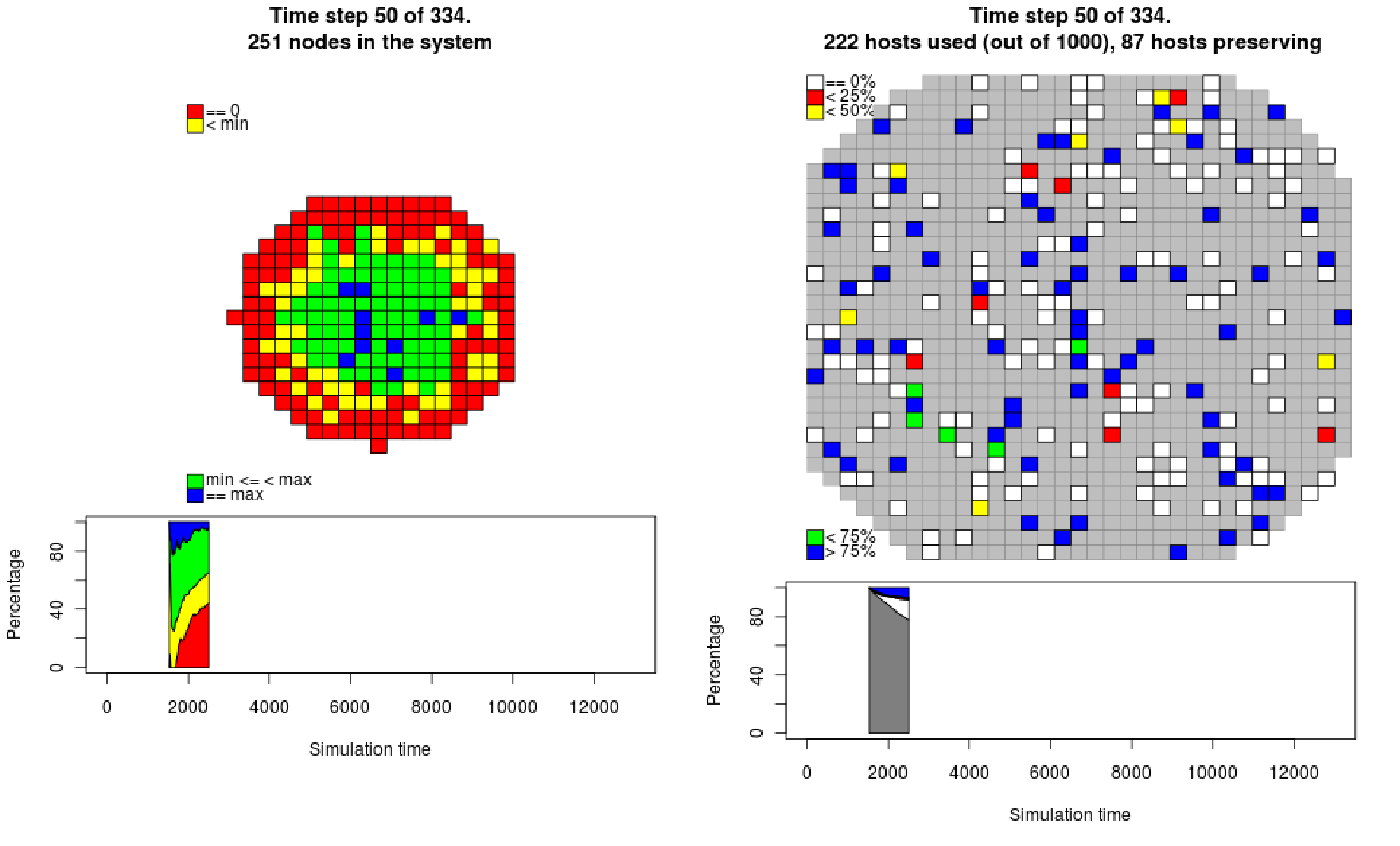}
\MySubfigure{\MySimulationTime = 3500.}{sub:fig:tl:4}{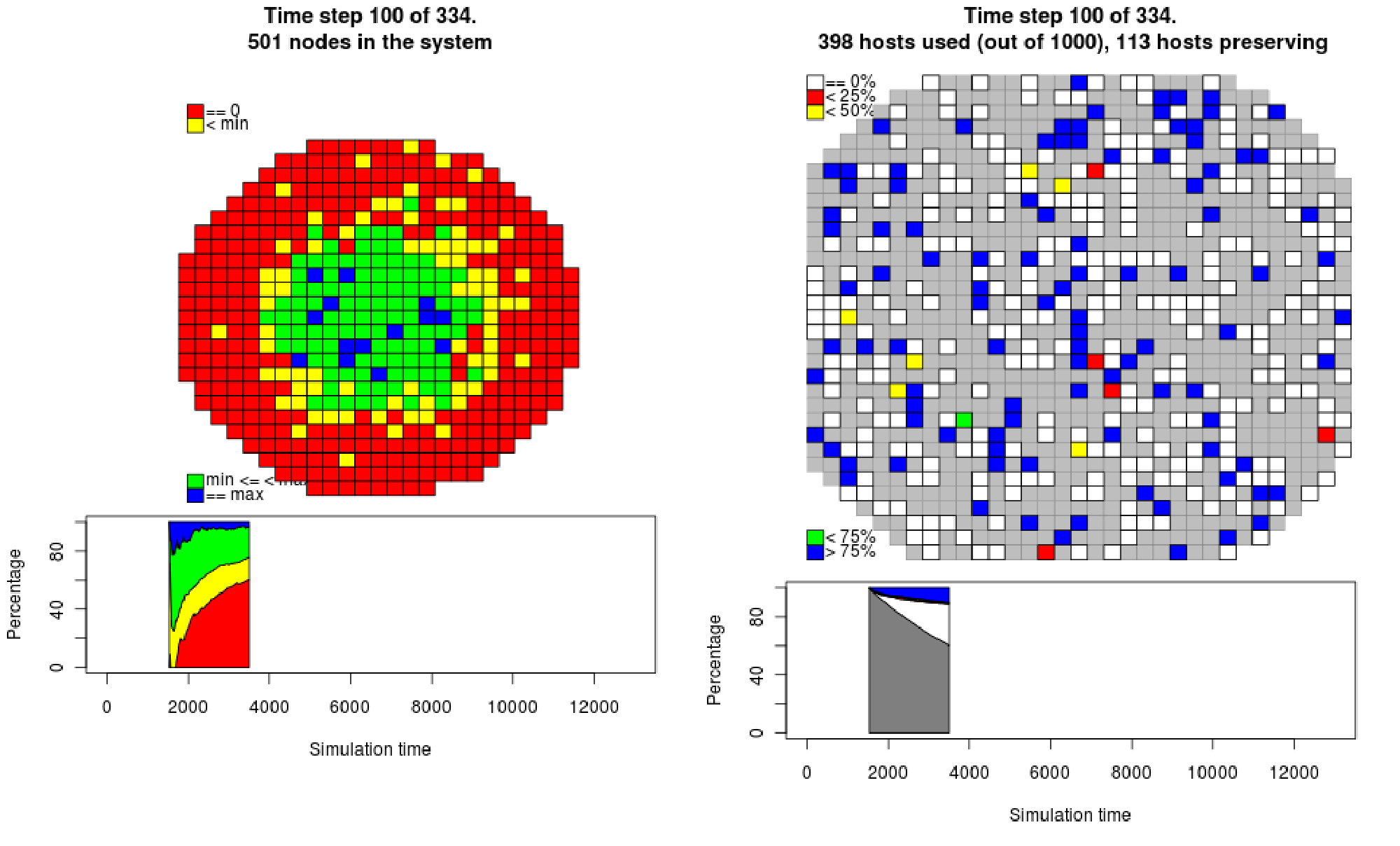}
 \MyCaption{The growth of a $\MyMaxDOs=500$ DO system captured at various time-steps.}
{The left half of each sub-figure shows the ``tree ring'' growth of the DO's portion
of the system.  The DO and host histograms
show the percentage of DO and hosts that are in their respective states as a function
of time.
  All DOs have been created and assigned to a host by $\MySimulationTime = 3500$.}
{fig:timestep:legend}
\end{figure*}

\MyIgnore{Detailed information about how preservation and host data are presented \MyFigureReference{fig:timestep:legend}.  }The initial DO
is plotted in the center of the left hand upper quadrant of each composite, Figure \ref{sub:fig:tl:1} shows the first 5 DOs in the system.
The one in the center is the oldest DO, while the others are younger.  The five DOs currently in the system 
live on hosts in the system.  Hosts can live anywhere on the network and where a particular host 
is drawn
immaterial.    The hosts in Figure \ref{sub:fig:tl:1} have a finite  capacity that their respective
system administrators have allocated to the preservation of copies of ``foreign'' DOs $\MyHostCapacity$.

At any point in time during the simulation, there will likely be a difference in the number
of preservation copies that the DOs want to create and the preservation capacity of all the hosts.  Reynolds' rules attempt to balance
these two requirements over time.  Figure \ref{sub:fig:tl:1} indicates that the DOs have each made some number of copies (they are colored yellow
vice red) and those copies are spread across some of the hosts in a non-even manner.  One host has used all its capacity (as shown in blue),
while one
has not used any (as shown in white).  The remaining hosts have used something in between those two extremes (they are yellow and red).
  In
Figure \ref{sub:fig:tl:1}, the histograms do not show too much information because of the initial internal simulation activity prior to
the introduction of the first DO.

In Figure \ref{sub:fig:tl:2}, the tree ring growth of the DOs is becoming more apparent.
  Older DOs have had more opportunities to make
preservation copies of themselves, therefore there is more green and blue in the center of the DO plot.  Many of the hosts are have reached
\MyHostCapacity, as indicated by the number of blue hosts.  The histograms are starting to become filled with data.  The DO histogram is starting
to show that the percentage of the DOs that have made some, but not all their preservation copies (those in yellow) is starting to grow,
while the percentage of those that have reached their goals is lessening.  The hosts histogram is starting to show that the percentage of the
hosts that have been discovered and added to the system (the grey area), is starting to decrease. A DO will be local to exactly one host.
A host may have more than one DO local to it.  A DO will not put a
preservation copy on any host that it lives on, or that already has a preservation copy of itself.

In Figure \ref{sub:fig:tl:3}, the tree ring presentation of the DO success at preservation is becoming more pronounced.
Younger DOs are struggling to make copies, while the old ones are maintaining their copies.  More of the hosts are being brought into
the system (the percentage of grey hosts is decreasing), but a significant percentage of the hosts are not being used for preservation
(those shown in white).

In Figure \ref{sub:fig:tl:4}, all DOs have been introduced into the system.  The tree ring preservation effect is still evident,
and some of the new DOs have been fortunate enough to make some number of preservation copies (as shown by the yellow markers in the sea
of red).  The percentage of hosts that are still not preserving any DOs is still significant, and the percentage of hosts that have reached
\MyHostCapacity~is holding constant.  The system will continue to evolve until it reaches a steady state, when those DOs that have
preserved as many copies of themselves as
they can based on their knowledge of hosts that have excess preservation capacity.  The system steady state
for this particular graph is shown in Figure \ref{sub:figure:1}.

\begin{figure*}
\newcommand{\MySubfigure}[3]{\subfigure[#1]{\label{#2}\includegraphics[width = 4.20in] {E-#3-last}}\\}
\centering
\MySubfigure{Least aggressive preservation policy. System stabilization at \MySimulationTime = 8195.}{sub:figure:1}{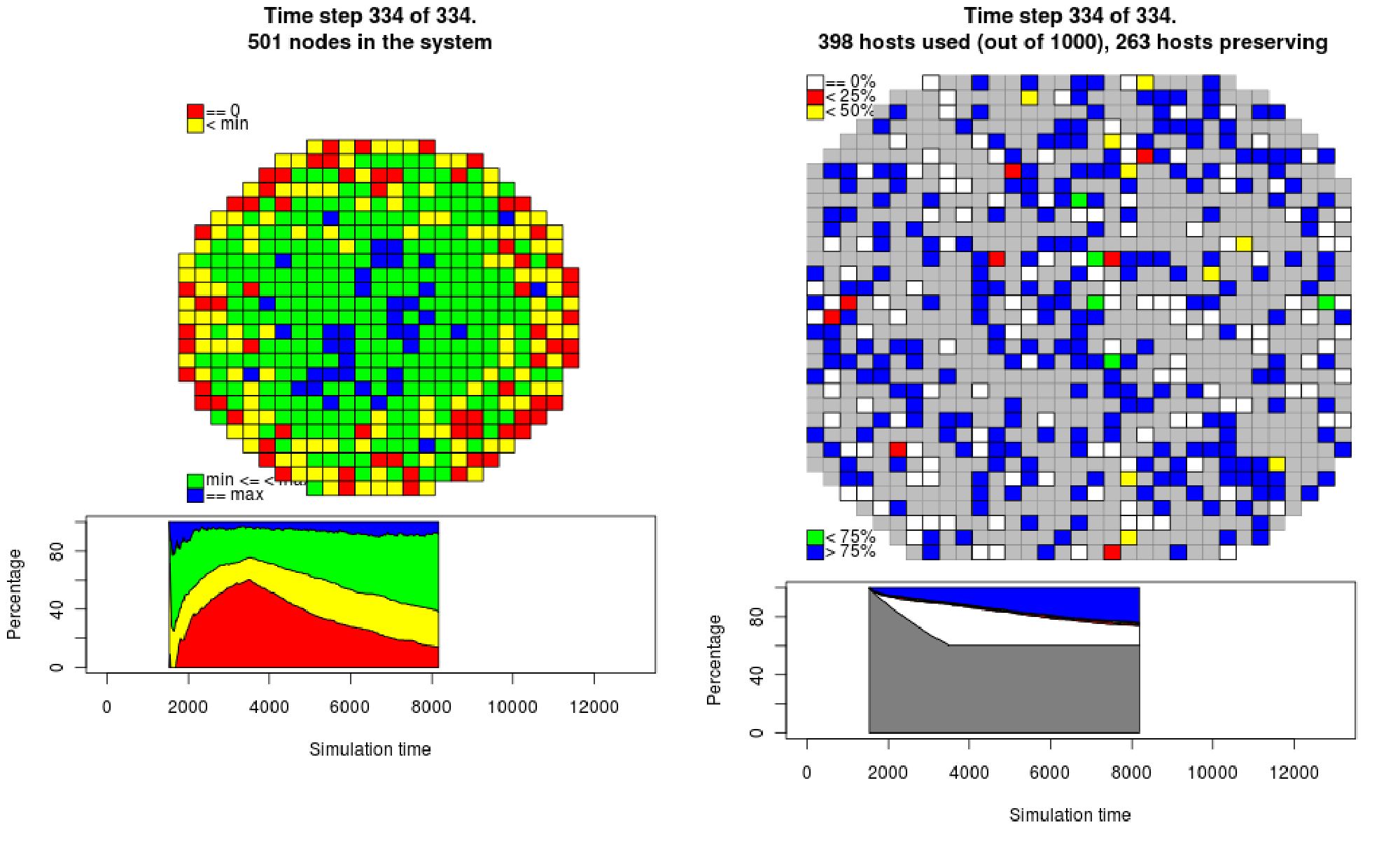}
\MySubfigure{Moderately aggressive preservation policy.  System stabilization \MySimulationTime = 12599.}{sub:figure:2}{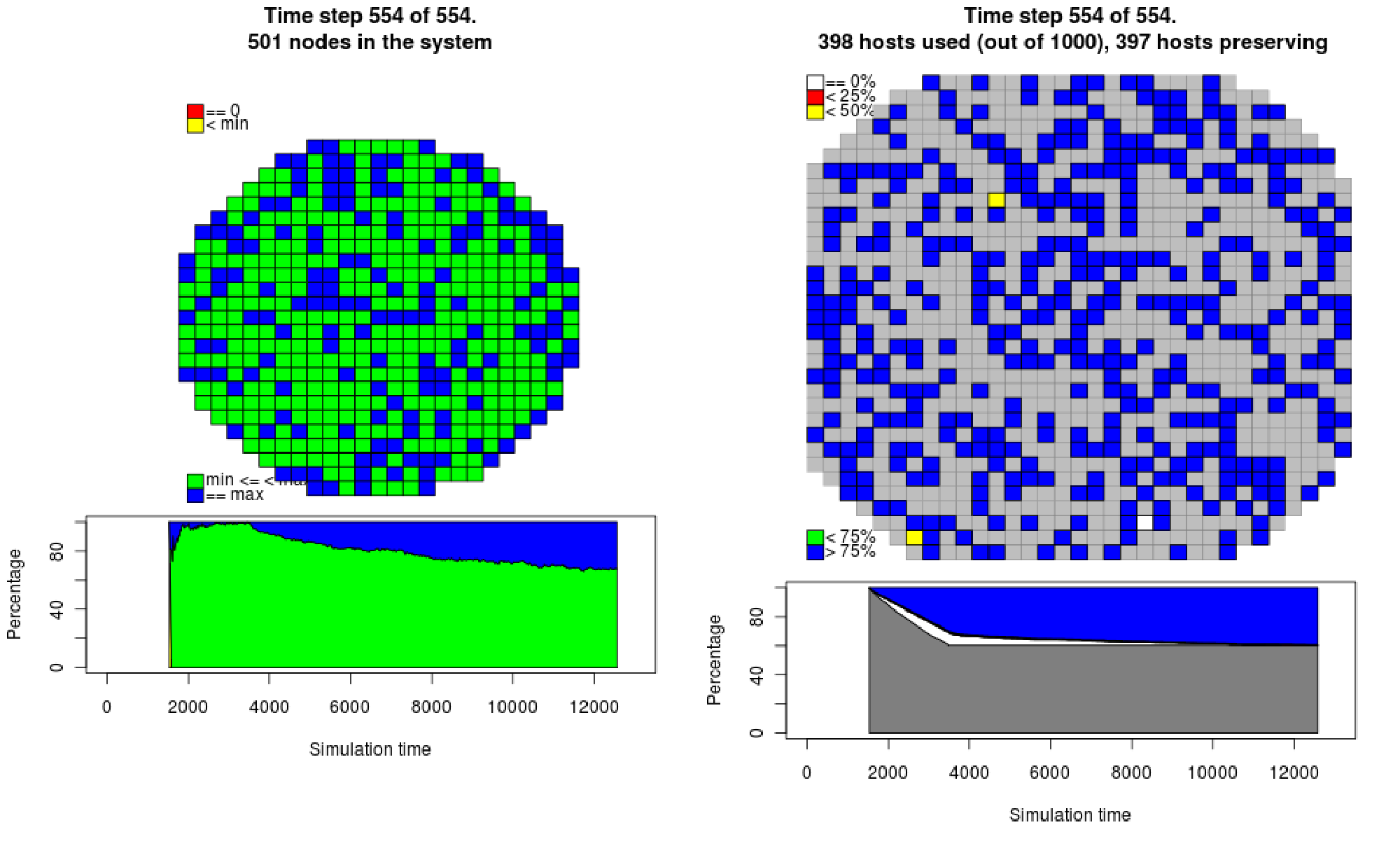}
\MySubfigure{Most aggressive preservation policy.  System stabilization at \MySimulationTime = 7521.}{sub:figure:3}{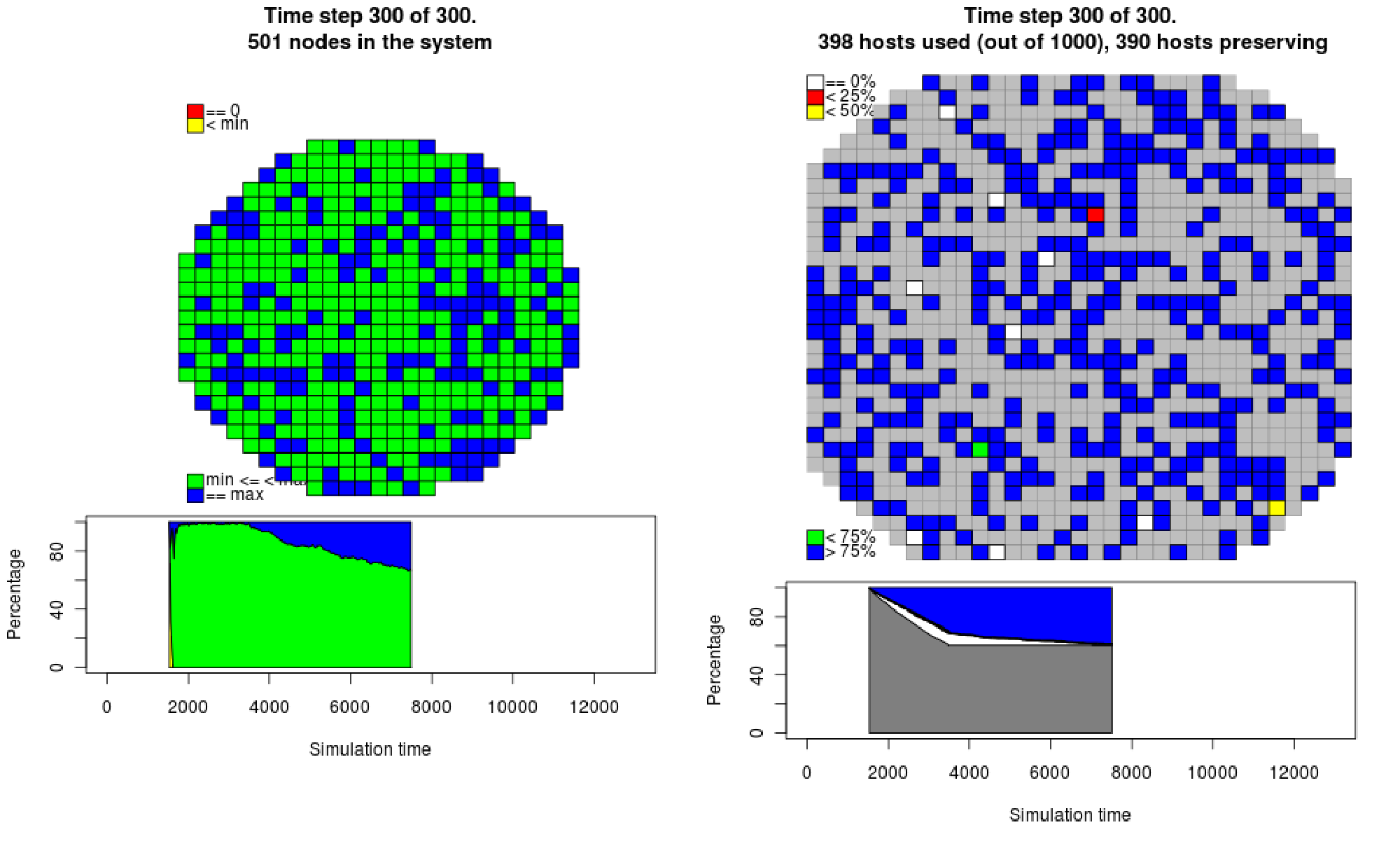}
\MyCaption{Time lapsed comparison of different preservation policies.}
{Using the Most aggressive policy results in a  higher percentage of DOs
meeting their preservation goals sooner and makes more efficient use
of limited host resources sooner.}
{fig:1}
\end{figure*}


\begin{figure*}
\newcommand{\MySubfigure}[3]{\subfigure[#1]{\label{#2}\includegraphics[width = 2.25in] {#3.eps}}}
\MySubfigure{DO$_{1,c,h}$, Least aggressive preservation policy.}{sub:figure:0:1}{0-1.png}
\MySubfigure{DO$_{250,c,h}$, Least aggressive preservation policy.}{sub:figure:250:1}{250-1.png}
\MySubfigure{Sum of all DOs, Least aggressive preservation policy.}{sub:figure:all:1}{all-1.png}
\\
\MySubfigure{DO$_{1,c,h}$, Moderately aggressive preservation policy.}{sub:figure:0:2}{0-2.png}
\MySubfigure{DO$_{250,c,h}$, Moderately aggressive preservation policy.}{sub:figure:250:2}{250-2.png}
\MySubfigure{Sum of all DOs, Moderately aggressive preservation policy.}{sub:figure:all:2}{all-2.png}
\\
\MySubfigure{DO$_{1,c,h}$, Most aggressive preservation policy.}{sub:figure:0:3}{0-3.png}
\MySubfigure{DO$_{250,c,h}$, Most aggressive preservation policy.}{sub:figure:250:3}{250-3.png}
\MySubfigure{Sum of all DOs, Most aggressive preservation policy.}{sub:figure:all:3}{all-3.png}

\MyCaption{Showing total messages sent and received by an early DO, a mid-simulation Do and all DOs.}
{The shape of the message sent curves (in black) for the early node is different based on the
preservation policy (see Figures \ref{sub:figure:0:1}, \ref{sub:figure:0:2} and \ref{sub:figure:0:3}).
While the shape of messages received curve (in red) remains almost the same.  This behavior is
contrasted with the mid-simulation node (see Figures \ref{sub:figure:250:1}, \ref{sub:figure:250:2} and \ref{sub:figure:250:3}).
  The mid-simulation node message sent curve is constant
regardless of the preservation policy.  The growth and maintenance phases are shown in light blue and light green respectively.
}
{fig:messages}
\end{figure*}

Figure \ref{fig:1} shows the steady state condition of the same system using
the three different preservation policies.  All DOs have been introduced into the system
by $\MySimulationTime=3500$ (as shown by the ``kink'' in the percentage of hosts that are used histogram).
Each preservation policy resulted in a significantly different time to reach a steady state.
 The hosts have enough preservation capacity
to accommodate the preservation needs of the DOs, a Boundary High condition \MyTableReference{tbl:systemTable}.
If the DO can locate
enough unique hosts via its friends, then it will be able to meet its preservation goals.  These
representative values for number of DOs, desired preservation levels and host preservation capacity
were chosen to illustrate the interaction between the DOs as they move preservation copies from 
one host to another while attempting to maximize the preservation needs of  most of the DOs.

The Least aggressive policy reaches steady state
at $\MySimulationTime=8195$  \MyFigureReference{sub:figure:1} and  a significant
percentage of the DOs have not been able to make any preservation copies (as shown
by the lower-most  (red) band in the histogram).  As shown in the node half of
the figure, many of the hosts are not preserving any DOs and those hosts that are preserving have
reached their capacity.

The Moderately aggressive policy reaches steady state at $\MySimulationTime=12599$ \MyFigureReference{sub:figure:2}.
Prior to $\MySimulationTime=3500$, most of the DOs have made most of their preservation copies. After that time,
the percentage achieving \MyFamilySizeMax~ slowly increases until the system reaches steady state.
The hosts' preservation capacity is used by the DOs in the system almost as
quickly as the hosts come on line.  This is indicated by the very narrow white region between
the unused host region and the totally used region.  At steady state, only a very few
of the hosts have not been totally used (as shown by the few host usage squares that are neither
blue or grey).

The Most aggressive policy reaches steady state after $\MySimulationTime=7521$ \MyFigureReference{sub:figure:3}.
Close examination of the host histograms in Figures \ref{sub:figure:2} and \ref{sub:figure:3}
 show almost identical
behavior both prior to $\MySimulationTime=3500$ and at steady state.  Comparing the host usage plot in the two figures show
that slightly more hosts have unused capacity based on a Most aggressive policy than a Moderately aggressive
policy (390 versus 397).  Based on \MyMaxDOs~DOs in the system, the difference between the two 
policies host under utilization is not  significant.

\subsection{Communications}
From the DO's perspective, there are two distinct phases of communication.  The first is when the
DO is \emph{wandering} through the graph and collecting information from DOs that are already connected
into the graph, this called the \emph{growth} phase.  The second is after the DO is connected into the graph 
and is called the \emph{maintenance} phase.  During the growth phase, the DO is actively communicating with other DOs.
While in the maintenance phase, the DO is responding to queries and communications from other DOs.  This
change in communication patterns occurs at  approximately $\MySimulationTime = 3500$ in Figure \ref{fig:messages}.
Figure \ref{fig:messages} shows the communications for 2 different DOs and the system in total as
a function of the preservation policy.
 DO$_{1,c,h}$ and DO$_{250,c,h}$ were chosen to represent the messaging profiles of all DOs
to see if the  profile changes as a function of when a DO enters the system.  Time in
Figure \ref{fig:messages} runs until $\MySimulationTime= 15000$
 and messages are counted in time bins sized to 100 simulation events.

Looking at figures \ref{sub:figure:0:1}, \ref{sub:figure:250:1}, \ref{sub:figure:0:2}, \ref{sub:figure:250:2},
\ref{sub:figure:0:3} and  \ref{sub:figure:250:3},  there is a
marked difference in the communication curves between DO$_{1,c,h}$ and DO$_{250,c,h}$.  These curves (with only
minor differences) are consistent across all preservation policies.  DO$_{1,c,h}$ (the earliest DO introduced
into the system), sends a rather modest number of messages $O(2n)$ to DOs that are also in the
system as DO$_{1,c,h}$ attempts to create preservation copies.  Under the least aggressive policy \MyFigureReference{sub:figure:0:1},
DO$_{1,c,h}$ sends a few messages per time bin until the system enters the maintenance phase.
The number of messages sent during the moderately aggressive policy is nominally the same  \MyFigureReference{sub:figure:0:2}.
While the Most aggressive policy results in messages for just a couple of time bins and then virtually no messages are sent  \
\MyFigureReference{sub:figure:0:3}.  Regardless of the preservation policy, the number of messages that DO$_{1,c,h}$ receives
is about the same.

Comparing the message curves for DO$_{1,c,h}$ and DO$_{250,c,h}$ indicates that the system discovered by the later DO
is very different than the one discovered by the earliest DO.  The late arriving node has more than enough opportunities to
satisfy its preservation goals when first introduced into the system.  DO$_{250,c,h}$ sends all of its messages in one
time bin and virtually nothing thereafter. This behavior is constant across all preservation policies and indicates
that the late arriving DOs are able to connect with another DO in very short order and
almost immediately enter into the maintenance phase of their existence.  The maintenance phase of the system
corresponds to a combination of the \textit{velocity matching} and \textit{flocking centering}.

The system is in a  growth phase from about $\MySimulationTime = 1500$ to $\MySimulationTime = 3500$
as shown by the rising curves in the ``Sum of all DOs'' sub-figures \ref{sub:figure:all:1}, \ref{sub:figure:all:2} and \ref{sub:figure:all:3}.  During the growth phase,
the \emph{wandering} node is sending and receiving a lot of messages while attempting to make its initial connection into
the graph.  After $\MySimulationTime = 3500$, the
system is in a maintenance phase when the system is attempting to balance the preservation needs of the DOs
with the capacity of the hosts. Comparing the messages curves for the entire system
 Figures \ref{sub:figure:all:1}, \ref{sub:figure:all:2} and \ref{sub:figure:all:3}
shows that there is no qualitative  difference between the number of messages sent and received in the system
based on preservation policy.  The nuances of the message curves for early DOs is lost as the
size of the system increases.

\begin{sloppypar}
\subsection{Messages Sent and Received as the System Grows in Size}
\end{sloppypar}

Figures \ref{fig:1} and  \ref{fig:messages} show the efficacy and communication costs associated
with a system with $\MyMaxDOs = 500$  and $\MyMaxHosts=1000$.  These values allowed the simulation to execute quickly, thereby
enabling more options and combinations to be investigated.  After determining that  at least a moderately aggressive
preservation policy enabled a high percentage of DOs to meet at least their \MyFamilySizeMin~
goals, the next area of investigation was to determine how the total number of messages
changes as a function of system size.  Figure  \ref{fig:messages} clearly shows that there 
 different types of communication during the growth and
maintenance phases.  During the maintenance phase, the DOs are attempting to spread their preservation copies
out across all the unique hosts in their friend's network.  A cost function was 
developed to quantitatively investigate
the performance of the various preservation policies focusing on the number of messages
sent and received.  Each  preservation status \MyFigureReference{fig:legend} was assigned a value
from 1 to 4 corresponding to the range 0 to \MyFamilySizeMax~ and scaled 0 to 1.  At
each  \MySimulationTime~ the cost performance of the system was evaluated and the 
number messages sent and received up to that point was summed.  The results are shown
in Figure \ref{fig:communications}.  The Most aggressive and moderately aggressive policies achieve approximately 
the same level of effectiveness, but the Most aggressive achieves that level
twice as fast as the moderately aggressive and only sending half as many messages.

The current simulation is a Boundary High condition \MyTableReference{tbl:systemTable}. One of the contributing factors to  spreading
preservation copies across many hosts 
is the limited capacity of the hosts to support preservation.  In order to remove the effects of maintenance communications and
focus purely on the effect of the number of DOs in the system, a series of simulations were
run using a Feast condition environment \MyTableReference{tbl:systemTable} where $\MyHostCapacity = 2 * \MyMaxDOs$.  This ensured that there
would be room on the host for any DO that discovered the host via  one of their friends.
\MyIgnore{Figure \ref{fig:messageTotals} shows number of messages exchanged in the system based on the number of DOs in the system and the
preservation policy.}
Based on the simulations, the total number of messages exchanged during the
growth phase approximates $O(n^{2})$ and the incremental messaging cost
of each new DO to the system is $O(2n)$.

\begin{figure}
\centering
\includegraphics[width=2.5in, angle=-90]{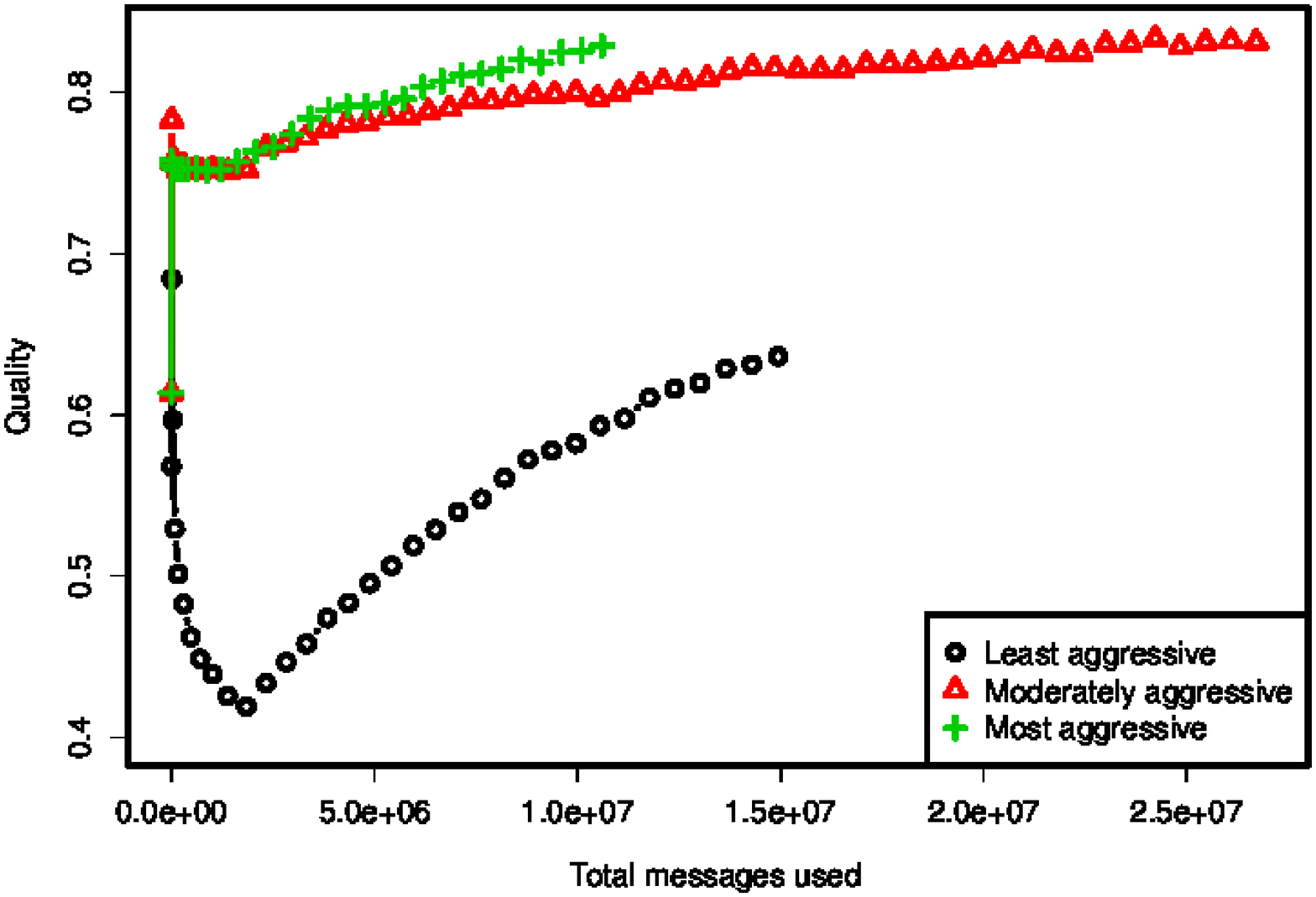}
\MyCaption{The preservation effectiveness as a function of policy and number of messages sent
and received.}
{}
{fig:communications}
\end{figure}

\section{Conclusion}
We have shown that implementing Reynolds' ``boid'' model with a limited number of rules
in an autonomic system can result
in digital objects (DOs) behaving in a manner that works towards the betterment
of the whole by occasionally sacrificing an individual.  Using simulations, we
investigated different policies that DOs could use when make preservation copies
of themselves.  Based on simulations of 500 DOs and 
hosts with limited preservation capacity; the Most aggressive preservation policy enabled
the DOs to attain the same preservation percentage in half the time as 
a Moderately aggressive policy  while exchanging only half 
as many messages.  An aggressive policy will try to make up to
\MyFamilySizeMax~ copies as it can at its first  opportunity and then single copies thereafter.
 A Moderately aggressive preservation policy  will try to make up to \MyFamilySizeMin~ copies 
at its first opportunity and then single copies thereafter until it reaches \MyFamilySizeMax.
The least aggressive preservation policy attempts to make 1 preservation copy per opportunity
until it reaches \MyFamilySizeMax.

There are two distinct communication message profiles; one prior to all the DOs being
introduced into the system and one after.  
The system's \emph{growth} period  is characterized by many messages
being sent from the \emph{wandering} DO and few being received while the DO attempts to
make its appropriate number of preservation copies.  The \emph{maintenance} period is
characterized by a relatively few number of messages as the DO is directed to sacrifice
preservation copies for the greater good of the graph, and subsequently having to create
copies anew.  There are distinct differences between the growth message profiles of
new and late arriving DOs, based on the preservation policy.
\begin{table*}
\centering
\scalebox{1.0}{
\begin{tabular}{|p{0.25\textwidth}|m{0.20\textwidth}|m{0.20\textwidth}|m{0.20\textwidth}|}
    \MyHline
    \textbf{Named host capacity}  &
    \multicolumn{3}{c|}{\textbf{Preservation policy}} \\
    \cline{2-4}
    (condition)&
    \multicolumn{1}{c|}{Least aggressive} & 
    \multicolumn{1}{c|}{Moderately aggressive} & 
    \multicolumn{1}{c|}{Most aggressive} \\
    \hline
    \textbf{Famine} \linebreak ($\MyFamilySizeMin \le \MyFamilySizeMax    < \MyHostCapacity$) \linebreak & \multirow{5}{0.20\textwidth}{Lowest percentage of DOs achieving preservation goal.} & \multicolumn{2}{c|}{
      \multirow{2}{0.20\textwidth}{Equally marginally effective.}
    } \\
    $\mbox{\textbf{Boundary Low}}$ \linebreak ($\MyHostCapacity = \MyFamilySizeMin \le \MyFamilySizeMax$)  \linebreak & & \multicolumn{2}{c|}{}\\
    \cline{1-1}
    \cline{3-4}
    \textbf{Straddle} \linebreak($\MyFamilySizeMin \le \MyHostCapacity \le \MyFamilySizeMax$) \linebreak &  & \multirow{3}{0.20\textwidth}{The baseline against which others are measured.}&\multirow{3}{0.20\textwidth}
    {\textbf{This preservation policy is twice as efficient as the Moderately aggressive policy for these named conditions.}}\\
    $\mbox{\textbf{Boundary High}}$ \linebreak($\MyFamilySizeMin  \le \MyFamilySizeMax  = \MyHostCapacity$) \linebreak & & &\\
    \textbf{Feast}  \linebreak($\MyHostCapacity < \MyFamilySizeMin \le \MyFamilySizeMax$) \linebreak & & &\\
    \MyHline
  \end{tabular}
}
\MyCaption{The effectiveness of various preservation policies based on named host capacity conditions.}  
{In this table we have taken the liberty to abuse the definitions of \MyHostCapacity, \MyFamilySizeMin~ and \MyFamilySizeMax~ by interpreting them to apply to the total system, vice a single host or DO. 
  In all cases, the Least aggressive
policy was the least successful at meeting the system's preservation goals.
Under the Famine and Boundary Low conditions, when it would be impossible to meet preservation goals,
both the 
Moderately and Most aggressive policies arrived at approximately the same steady state
situations after exchanging approximately the same number of messages. Straddle conditions 
would permit some DOs to achieve their goals, if the DOs were fortunate.  Straddle 
results under Moderately and Most aggressive policies are comparable and the
Most aggressive reaching steady state after exchanging about $\frac{1}{2}$ as many messages
as the Moderately aggressive policy.
Boundary High and Feast conditions have enough capacity for the system to meet its preservation 
needs, and the Most aggressive policy operates about twice as efficiently as the Moderately 
aggressive policy.}
{tbl:systemTable}
\end{table*}

\section{Acknowledgment}\label{sec:acknowledgement}
 This work supported in part by the NSF, Project 370161.

\bibliographystyle{abbrv}
\bibliography{master}
\end{document}